\documentclass[manuscript]{aastex63}

\shorttitle{Lensed quad}
\shortauthors{Jones et al.}

\begin{document}

\title{Image Flux Ratios of Gravitationally Lensed HS 0810+2554 with High Resolution Infrared Imaging}

\author{Terry Jay Jones, Liliya L. R. Williams}
\affil{Minnesota Institute for Astrophysics, School of Physics and Astronomy\\116 Church St SE, University of Minnesota, Minneapolis, MN 55455, USA}

\author{Steve Ertel}
\affil{Steward Observatory, Department of Astronomy, University of Arizona, 993 N. Cherry Ave, Tucson, AZ 85721, USA}
\affil{Large Binocular Telescope Observatory, 933 Cherry Avenue, Tucson, AZ 85721, USA} 

\author{Philip M. Hinz}
\affil{Steward Observatory, Department of Astronomy, University of Arizona, 993 N. Cherry Ave, Tucson, AZ 85721, USA}
\affil{Center for Adaptive Optics, UC Santa Cruz, 1156 High Street, Santa Cruz, CA 95064}

\author{Amali Vaz}
\affil{Steward Observatory, Department of Astronomy, University of Arizona, 993 N. Cherry Ave, Tucson, AZ 85721, USA}

\author{Shane Walsh}
\affil{Large Binocular Telescope Observatory, 933 Cherry Avenue, Tucson, AZ 85721, USA}

\author{Ryan Webster}
\affil{Steward Observatory, Department of Astronomy, University of Arizona, 993 N. Cherry Ave, Tucson, AZ 85721, USA}

\begin{abstract}

We report near simultaneous imaging using LMIRCam on the LBTI of the quadruply imaged lensed quasar HS 0810+2554 at wavelengths of 2.16, 3.7 and $4.78~\micron$ with a Full Width Half Max (FWHM) spatial resolution of $0\farcs13$, $0\farcs12$ and $0\farcs15$ respectively, comparable to HST optical imaging. In the $\rm{z} = 1.5$ rest frame of the quasar, the observed wavelengths correspond to 0.86, 1.48, and $1.91~\micron$ respectively. The two brightest images in the quad, A and B, are clearly resolved from each other with a separation of $0.187 \arcsec$. The flux ratio of these two images (A/B) trends from 1.79 to 1.23 from 2.16 to $4.78~\micron$. The trend in flux ratio is consistent with the $2.16~\micron$ flux originating from a small sized accretion disk in the quasar that experiences only microlensing. The excess flux above the contribution from the accretion disk at the two longer wavelengths originates from a larger sized region that experiences no microlensing. A simple model employing multiplicative factors for image B due to stellar microlensing $(m)$ and sub-structure millilensing $(M)$ is presented. The result is tightly constrained to the product $m\times M=1.79$. Given the observational errors, the 60\% probability contour for this product stretches from $m= 2.6$, $M = 0.69$ to $m= 1.79$, $M = 1.0$, where the later is consistent with microlensing only.

\end{abstract}


\section{Introduction}

High resolution N-body simulations of the cold dark matter (CDM) structure formation of
the universe predicts that there should be hundreds of dark matter sub-halos with mass
$\sim 10^4 - 10^9{\rm{M}}_\odot$ within a massive halo $(\sim 10^{12}{\rm{M}}_\odot)$. The fact that only a few dozen dwarf galaxies have been observed within our own Milky Way \citep{mcco12}  has led to the so called ''missing
satellite'' problem \citep{klyp99, moor99, sawa14}. The inclusion of baryonic physics \citep{garr19, bull17, broo14, buck18, dutt16}  in models of the formation of galaxies has made progress on resolving this issue, but many aspects of dark matter substructure are still uncertain \citep{desp17, gome18, desp19, hsue19}.

While these sub-halos would remain undetected due to their lack of visible matter, they should still be
detectable through strong gravitational lensing \citep[e.g.][]{koch04}. Gravitational lensing provides a unique method for detecting the presence or absence of substructure in a galaxy outside of the Local Group. Any perturbation in the lens potential, be it stars or dark matter substructure, will create perturbations in the observed flux from some of the lensed images, if it is sufficiently compact and close to the images. It has been shown by \cite{mao98}, \cite{metc01}, and \cite{chib02} that the presence of substructure in the lens mass model can reproduce the observed flux ratios in multiply imaged lensed quasars, something that
smooth lens models can not do. 

\cite{hsue19} use a sample of seven quadruply imaged lensing systems and assuming a CDM cosmology and contribution from low mass halos along the line-of-sight, they infer an average total mass fraction in substructure that is in rough agreement with the predictions from CDM hydrodynamical simulations. Their result is significantly different when compared to previous studies that did not include line-of-sight haloes \citep{xu15}. Their data set is dominated by images with good flux ratios at radio wavelengths, but with less contribution to the analysis from Mid-Infrared (MIR) observations which probe smaller scales. Several lensed systems, including HS 0810+2554, have images that are probably resolved at radio wavelengths and were dropped from their initial sample of 14. \cite{nier19} report observations of 8 quads, separating out emission from the quasar narrow line emission region (which is not subject to microlensing) and conclude that simple, smooth halo mass distributions are inadequate, indicating the need for more complex mass distributions. Although HS 0810 was in their sample, the small angular distance between images A and B made the analysis by \cite{nier19} difficult, and narrow line flux ratios were not reported. 

\cite{dobl06} point out that the angular size of
the source and the size of the perturber must be on the same order of magnitude for the
source to be appreciably magnified.  MIR observations have the advantage that in the rest frame of the lensed quasar, they span a range in wavelengths that are sensitive to the emission of both the very small sized accretion disk in the quasar which is subject to the effects of both stellar microlensing and substructure millilensing, and the much larger dusty torus which is subject only to millilensing \citep{slus13}. Therefore, if we compare the flux ratios in a lensed quasar at different wavelengths, corresponding to different source sizes, we can quantify the size and location of substructure in the lens galaxy. This approach was taken by \cite{fade11, fade12} using K and L’ bands for six lenses with image separations $> 1\arcsec$
and by \cite{chib05} at $11\micron$ for two lenses with image separations $\sim 1.0$ and $0.5\arcsec$. The closer the two brightest images in a quad are together, the less influence larger scale variations in the halo potential (e.g. ellipticity) and the effects of time delay will have \citep{lemo17}. Thus, quad images with very close separation are preferred in the search for evidence of halo substructure.

In this paper we present observations of the quad lensed system HS 0810+2554 \citep{reim02, char14, jack15, nier19} at wavelengths from $2-5\micron$. The background quasar is at a redshift of ${\rm{z}}=1.5$ but the lensing galaxy is at an unknown redshift. The two brightest images are separated by only $0.19\arcsec$ and at a redshift of z = 1.5, the observations span a range in rest frame wavelength of $0.8-2.0~\micron$. Over this wavelength range the flux arises purely from the accretion disk at the shorter wavelength, and from a combination of accretion disk and dusty torus at the two longer wavelengths \citep[e.g.][]{koba93, suga06}.

\section{OBSERVATIONS}\label{sec-obs}

We observed HS 0810 on UT 2019 February 24 with LMIRCam \citep{skru10} on the LBTI \citep{hinz16} using a single 8.4 m primary mirror of the Large Binocular Telescope (LBT).  The focal plane scale was $0.0108 \arcsec$ per pixel. Images were made in the Ks $(2.16 ~\micron)$, L' $(3.7 ~\micron)$ and M $(4.78 ~\micron)$ filters and the natural seeing was $0.8\arcsec$. We also made a test observation along with a standard star on UT 2017 February 17 in the L' filter (only) with natural seeing of $1.2\arcsec$. Filters and integration times are listed in Table 1. The AO secondary was operating in natural guide star mode for both epochs and had to lock onto the optical images of the two adjacent bright images in the quad together. This probably explains the slightly elongated images we obtained for the quad compared to the standard star.

\begin{deluxetable}{ccccc}
\tablecaption{LMIRCam Observations}
\tablewidth{0pt}
\tablehead{ \colhead{Date} & \colhead{Filter} & \colhead{frame time (ms)} & \colhead{Int Time (sec)} & \colhead{Total Int Time (min)}}
\startdata
2019 & Ks $(2.16~ \micron)$  & 1475 & 30.1 & 30.1 \\
2019 & L' $(3.7~ \micron)$ & 13.7 & 1.0 & 26.7  \\
2019 & M $(4.78~ \micron)$ & 13.7 & 0.412 & 32.9 \\
2017 & L' $(3.7~ \micron)$ & 290 & 0.874 & 9.6 \\ 
\enddata
\end{deluxetable}

For the 2017 observations in the L' filter, several dither positions on the array were used, but most often with a separation of $4.5\arcsec$. In 2019 we imaged HS 0810 at two dither positions on the array, separated by $8\arcsec$. Images at each dither position in the 2019 observations were taken in sets of 10, 100 and 200 integrations at Ks, L', and M respectively. Slightly elongated images, probably due to the AO locking onto the blended image, were evident. The axis of elongation rotates between wavelengths due to the rotation of the sky on the focal plane over the length of time the observations were made. The images occupied a small portion of the array, minimizing errors in flat fielding. Simple aperture photometry was used to determine the brightness of the well isolated sources C and D in the image, whereas an elliptical point spread function (PSF) was used to separate the slightly blended images of A and B, as explained in Section 3. Formal statistical errors were computed using the measured rms fluctuations in the background, which was determined to be flat within the vicinity of sources C and D. For blended sources A and B, a small systematic error of 0.01 mag. was added to account for any potential residual uncertainties in the background level (see Section 3) due to the more complicated PSF fitting.

Due to time and weather constraints in 2019, we were unable to observe standard stars on the same night as HS 0810 to fix the fluxes at each wavelength on a photometric scale. We have examined standard star observations on previous nights at Ks, L' and M and found that the relative response of the detector between filters, adjusted for airmass and stellar colors, is consistent to within a few percent. We used this information to build the relative fluxes between filters for our observations of HS 0810. In this way we can place the fluxes of the images in HS 0810 on a spectral energy distribution (with a flux normalization) with confidence. This does not affect the relative brightness between images at each wavelength which are, within the photometric and systematic errors, very well determined.

\section{Analysis}

Gray scale images of a star (HD 74721) and HS 0810 from the 2017 test observations are shown in Figure 1. The stellar image has had dithers subtracted, but has not been processed for bad pixel and pattern removal nor has the image been flat fielded. The image of HS 0810 has been fully processed. The individual lensed images are labeled according to \cite{reim02}. A radial profile of the stellar image is shown in Figure 2, where the measured FWHM is $0.10\arcsec$. 

Gray scale images of the quad from the 2019 observations are shown in Figure 3. The lensing galaxy itself is barely visible in the Ks band, and finding its centroid is impossible using our data. Consequently we can not measure the positions of the lensed images with respect to the lensing galaxy. The positions relative to image A are, within an error of $\pm 0.003\arcsec$, identical to the values listed in the CfA-Arizona Space Telescope LEns Survey (CASTLES) website \citep{muno98}. The brightness for images C and D were computed using standard aperture photometry on the images. For images A and B, their individual point spread functions (PSF) overlap. For these two images, the PSF was modeled with a simple two dimensional Voigt profile with a slight ellipticity. The Gaussian width, dampening wings parameter, and orientation (long axis) were derived from image C at each wavelength. The measured values for the Gaussian width ($1\sigma$) were $0.055\arcsec$, $0.051\arcsec$, and $0.064\arcsec$ at Ks, L', and M respectively. The dampening wings made a small, but detectable contribution to the overall PSF. The ellipticity was 15\% at Ks and L', and 5\% at M. 

A synthetic image of the blend was then formed using two model Voigt profiles with adjustable spacing and peak intensity. This model was subtracted from the image and the spacing and peak intensity of each profile adjusted until the residual intensity was less than 1\% of the total intensity of the A+B combination. A cut through the centers of images A and B in the L' filter for the 2019 observations is shown in Figure 4 along with the model fit. A slight tilt to the baseline background can be seen in the image cut that is not modeled. Since our PSF modeling is more complicated that the simple aperature photometry used for images C and D, a systematic error of 0.01 mag. was added to the error budget to account for uncertainties in method. The rest of the error budget was determined by comparing separate sets of observations that were co-added to make the final images and computing the statistical variation between these  subsets. The results are given in Table 2, with the ratios of images B, C and D to A expressed as magnitude differences, e.g., for images B and A, $\Delta m =  - 2.5\log ({\rm{B/A}})$.

\begin{deluxetable}{cccccc}
\tablecaption{Observed and Model Flux Ratios Expressed As Magnitude Differences}
\tablewidth{0pt}
\tablehead{\colhead{Date} & \colhead{Filter} & \colhead{$\Delta$m(B/A)} & \colhead{Model $\Delta$m (sec. 4)} & \colhead{$\Delta$m(C/A)} & \colhead{$\Delta$m(D/A)}}
\startdata
2019 & $2.16 ~\micron$  & $0.635\pm0.02$  & 0.638 & $1.60\pm0.03$ & $2.75\pm0.05$ \\
2019 & $3.7 ~\micron$ & $0.339\pm0.02$ & 0.358 & $1.56\pm0.03$ & $2.45\pm0.05$  \\
2019 & $4.78 ~\micron$ & $0.224\pm0.03$ & 0.223 & $1.48\pm0.12$ & $2.39\pm0.22$ \\
2017 & $3.7~\micron$ & $0.40\pm0.03$ & -- & $1.42\pm0.04$ & $2.22\pm0.06$ \\
\enddata
\end{deluxetable} 

\begin{deluxetable}{ccccc}
\tablecaption{Observed Flux Ratio Comparison}
\tablewidth{0pt}
\tablehead{\colhead{Band} & \colhead{Year} & \colhead{$\Delta$m(B/A)} &  \colhead{$\Delta$m(C/A)} & \colhead{$\Delta$m(D/A)}}
\startdata
0.2--10 keV & 2002 & 0.20 & 0.73 & 2.3 \\
0.2--10 keV & 2013 & 0.35 & 0.26 & 1.8 \\
V & -- & 0.78 & 1.16 & 2.07 \\
$0.72~\micron$ & 2001 & 0.7 & 1.4 & 2.8 \\
I & -- & 0.85 & 1.66 & 2.98 \\
H & -- & 0.61 & 1.44 & 2.76 \\
Ks  & 2019 & 0.635   & 1.60 & 2.75 \\
L' & 2019 &  0.339 & 1.56 & 2.45  \\
L' & 2017 & 0.40 &  1.42 & 2.22 \\
M & 2019 & 0.224  & 1.48 & 2.39 \\
20 cm & 2014 & 0.02 & 0.37 & 0.59 \\
\enddata
\end{deluxetable} 

Figure 5 shows images A and B with scaled contours overlaying the gray scale image at each wavelength. In all 3 filters image A, which is the minimum in the lensing arrival time surface, is brighter than image B, which is the saddle point. This is consistent with these images being affected by substructure--stars or subhalos, where the saddle is expected to be demagnified with respect to the neighboring minimum \citep{sche02, saha11}. Clearly visible is the trend towards more equal brightness with increasing wavelength. Our goal is to determine if the ratio of B/A as a function of wavelength can be used to determine if there is substructure in the dark matter halo of the lensing galaxy or if all of the B/A ratios are consistent with microlensing only. Because the two images are very close on the sky in comparison to the scale of the quad system, their light paths are traveling through nearly identical regions in the gravitational potential of the lensing galaxy. If there were no microlensing and the dark matter halo associated with the lensing galaxy was smooth azimuthally in the region of images A and B, images A and B should have identical fluxes. 

We estimated the time delay between images A and B using a free-form lens modeling software PixeLens \citep{will04}, which is dependent on the (unknown) redshift of the lensing galaxy. \cite{mosq11} made an estimate of $z=0.89$ for the lensing system based on basic photometric properties of the galaxy. For $z=0.3$ ($z=0.6$, $z=0.89$), the time delay is $\sim 0.05$ days ($\sim 0.15$ days, $\sim 0.33$ days) respectively. These delays are much shorter than typical intrinsic quasar variability \citep{macl12}, and comparable to the $\sim$3 hours of telescope time we spent making the observations. For all practical purposes, our observations of the A and B quasar images can be considered simultaneous in the rest frame. 

Our flux ratios, expressed as magnitude differences, are also listed in Table 3, which includes ratios at other wavelengths from \cite{char16}, \cite{jack15} and \cite{reim02}, along the the epoch of observation. We have been unable to find information on when the CASTLES observations cited in \cite{char16} were made, so no epoch is given in the Table. Note that there is clear variation in the short wavelength (X-Ray -- Ks band) B/A ratio over time, ranging from $\Delta{\rm{m}} = 0.2$ in the X-Rays in 2002, to 0.635 at Ks in 2019. According to \cite{mosq11}, a typical timescale for a microlensing event in the HS 0810 system due to crossing the Einstein radius of a single star is $\sim24$ yrs, a timescale comparable to the elapsed time between the earliest and most recent observations. For a typical projected density of stars in the lensing galaxy, microlensing will likely result from the combined effect of many stars along the line-of-sight, rather than a single star. Modeling by \cite{wamb90} shows that microlensing is a non-linear process, and both a longer term trend in magnification in addition to variations on time scales shorter than the crossing time, combine to create the light curve of an image. This is similar to what we observe for the B/A ratio, which is always in the same sense $({\rm{A}}>{\rm{B}})$, but with significant variations over the last 20 years.

To establish the range in mass and size of halo sub-structure our observations are sensitive to, we list the relevant $source$ plane sizes in Table 4. The values for the stellar Einstein radius R$_{\rm{E}}$, the accretion disk radius R$_{\rm{AD}}$, and the radius of the broad line region R$_{\rm{BLR}}$, were taken from \cite{mosq11}. We estimated the size of the outer radius of the dusty torus from Table 1 in \cite{slus13} assuming HS 0810 has a luminosity of $L\sim 5\times 10^{45} - {\rm{ergs}} - {{\rm{s}}^{ - 1}}$ and that the torus is represented by the 'extended' model. The luminosity estimate was made using the black hole mass given in \cite{mosq11} and the mass -- luminosity relation in \cite{woo02}. We chose the extended torus model since the infrared flux from HS 0810 is greater than for the typical quasar (Figure 6), suggesting a more extensive emitting region. 

\begin{deluxetable}{cc}
\tablecaption{Source Plane Sizes}
\tablewidth{0pt}
\tablehead{\colhead{Source} & \colhead{Size (pc)} }
\startdata
R$_{\rm{E}}$ (stellar) & 0.006 \\
R$_{\rm{AD}}$ & 0.0003 \\
R$_{\rm{BLR}}$ & 0.03 \\
R$_{\rm{IR}}$ (outer) & 1.0 \\
\enddata
\end{deluxetable} 

With these size estimates we can roughly estimate the mass range for any halo sub-structure that can significantly influence the flux ratios we measure by using a Sersic radial profile to compute an Einstein radius. Using a suite of numerical simulations and fitting radial profiles with an Einasto function, \cite{nava04} derive a shape parameter of $\sim 0.17$ that fits the simulated haloes well. Using Figure 8 in \cite{dhar10}, this approximately corresponds to a Sersic profile index of $m_S\sim 6$. We use a Sersic profile because the Einstein radius can be analytically calculated for a given index and half-mass radius ${{\rm{R}}_{1/2}}$. For a mass of $10^4$M$_\odot$ and assuming the lensing galaxy is at z = 0.89, the Einstein radius ranges from 1.1 pc for ${{\rm{R}}_{1/2}}=0$ (point mass) to 0.5 pc for ${{\rm{R}}_{1/2}}=2$ pc projected onto the source plane. These sizes are comparable to the size of the infrared torus and represents a rough lower limit to the mass we can detect. The distance between images A and B in the lensing plane is $\sim 1.45$ kpc, again assuming the lensing galaxy is at z = 0.89. This corresponds to an Einstein radius for a point mass of $2.1\times 10^{10}$M$_\odot$, or $1\times 10^{10}$M$_\odot$ for ${{\rm{R}}_{1/2}}=2$ kpc. This represents a rough upper limit to the mass that can influence our observations.

The mean spectral energy distribution (SED) for quasars has been computed by \cite{pado17} and \cite{kraw15}. They find that the combination of a small, sub-parsec sized accretion disk with a simple power law SED well matches the rest frame optical flux. At longer wavelengths, starting at about $1.2~\micron$, flux from a much larger (few parsecs) dusty torus begins to contribute and eventually dominates the SED. \cite{kraw15} found a mean power law index for the accretion disk of F$_\nu \propto \nu^{-0.4}$. The dust in the torus is at a temperature of $\sim 1500$K \citep[e.g.][]{koba93, suga06}, and the flux from the dusty torus drops very rapidly shortward of $1.2~\micron$, which is on the Wien side of the emission spectrum for the dust. Our 3.7 and $2.16~\micron$ filters correspond to wavelengths of 1.48 and $0.86~\micron$ in the rest frame of the lensed quasar. At these rest wavelengths, the flux from a ${\rm{T}} = 1500$K black body in $\nu {\rm{F}}_\nu$ at the shorter wavelength is a factor of 12 lower than at the longer wavelength. Given that the dust emission in our $3.7~\micron$ filter is only about 40\% (see below) of the total, the contribution to our $2.16~\micron$ filter from hot dust emission is only $\sim 3$\%, and can be ignored.

We plot the SED for the entire (within a single photometric aperture) HS 0810 quad in Figure 6, along with the mean SED for quasars taken from \cite{kraw15}. The flux values plotted for HS 0810 were taken from the WISE, 2MASS \citep{skru06}, UKDISS \citep{hamb08} catalogs, and a quasar catalog by \cite{souc15}. As noted above, the K filter is likely dominated by the small accretion disk, and consequently could show considerable variation in flux over time due to stellar microlensing.  For this reason we plot the observed J and K band fluxes as a vertical bar, indicating the range of fluxes  between these data sets. For each of 2MASS and UKIDSS surveys, the J, H and K observations were made at the same time, so the J--K colors should represent the quasar accretion disk continuum at the epoch of observation. The 2MASS and UKIDSS H filter observations are strongly contaminated by H$\alpha$ and do not represent the continuum. The J band observations suggest a somewhat steeper (bluer) power law index of -0.3, but this filter could be contaminated by [OIII], although the effect should be small. 

We only have relative flux levels between our infrared filters and we have no simultaneous red and optical photometry. This makes it difficult to empirically determine the power law slope for the accretion disk in HS 0810 at the epoch of our observations. Based on the J--K colors from 2MASS and UKIDSS, we will fix the power law slope at -0.3 as shown in Figure 7. We have vertically scaled our relative flux measurements for the A+B pair to match the trend between the WISE W1 and W2 filters. Flux from images A and B by far dominate our images and would be expected to dominate the overall SED, and we will assume the $\bf{shape}$ of the overall SED represents the $\bf{shape}$ of the SED for the combination of images A and B. With this normalization, our Ks ($2.16~\micron$) point fits within the lower range of previous K band observations, although this is not a given. 

\section{Lensing Model}

In Figure 7 we plot the SED for HS 0810 where the dashed blue line shows a simple accretion disk power law with ${{\rm{F}}_\nu } \propto {\nu ^{-0.3}}$ that passes through our Ks flux. Working with this figure, we developed a simple model for the flux from images A and B by making the following assumptions. 1) The Ks flux is due entirely to a small sized (fraction of a parsec) accretion disk that can suffer microlensing in the lensing galaxy. 2) The longer wavelength flux consists of a contribution from the accretion disk (extrapolated from the Ks flux) and a contribution from a larger dusty torus that does not experience any microlensing. Expressed mathematically, we have:

\begin{eqnarray}
{I_{\lambda} } &=&  {I_{{{\rm{B}}_\lambda }}} + {I_{{{\rm{A}}_\lambda }}} \nonumber \\
{I_{{{\rm{B}}_\lambda }}} &=& {I_{{{\rm{B}}_{\lambda {\rm{AD}}}}}} + {I_{{{\rm{B}}_{\lambda {\rm{IR}}}}}}  \\
{I_{{{\rm{A}}_\lambda }}} &=& {MmI_{{{\rm{B}}_{\lambda {\rm{AD}}}}}} + {MI_{{{\rm{B}}_{\lambda {\rm{IR}}}}}} \nonumber
\end{eqnarray}

$I_{\lambda}$ is the total flux from A and B, AD indicates flux from the accretion disk alone, and IR represents the flux from the torus alone, as indicated in Figure 7. The factor $m$ is the ratio of $I_{\rm{A}}/I_{\rm{B}}$ due to microlensing. If lensing due to substructure in the dark matter halo was present in addition to microlensing, an additional magnification factor $M$ must be added. This factor, which we will call millilensing, effects both the emission from the accretion disk and the infrared emission from the torus. Observations of HS 0810 at radio wavelengths \citep{jack15} and in optical narrow emission lines \citep{nier19}, neither of which should be subject to microlensing, are consistent with no millilensing. For this reason we  initially set this value to $M=1.0$. Given the observed SED and our assumptions listed above, we can solve for the factor $m$ that provides a best fit to the data at all wavelengths and compare to the observations, where:

\begin{eqnarray} 
{R_\lambda } &=& {\left( {\frac{{{I_{{\rm{IR}}}}}}{{{I_{{\rm{AD}}}}}}} \right)_\lambda } \\
\frac{{{I_{{{\rm{B}}_\lambda }}}}}{{{I_{{{\rm{A}}_\lambda }}}}} &=& \frac{{1 + M + {R_\lambda }\left( {1 + mM} \right)}}{{mM(1 + M) + {R_\lambda }(1 + mM)}} \nonumber
\end{eqnarray}

The results of our model fit for the case with only microlensing, no millilensing $(M = 1)$, are shown in Figure 8, where we have expressed the flux ratios as a magnitude difference for comparison with the data (section 3). The data from Table 2 are plotted as points and the model fit corresponding to $m = 1.79$ is plotted as a solid line. The predicted magnitude differences from the model are listed in Table 2 along with the observed differences. 

Our simple model fits the data well, and are consistent with A/B ratios at radio wavelengths \citep{jack15} of 1.02+/- 0.06 and model results using narrow line emission \citep{nier19} that imply a flux ratio of 1.0 +/- 0.07. If substructure were present at a level that could significantly influence the observed flux ratios, then the measured ratios for B/A at the two longer wavelengths would lie above or below the simple model that is compatible with the flux ratio at Ks. In this case, the $\bf{product}$ $m\times M$ determines the flux ratio at Ks and the inclusion of the millilensing factor changes the flux ratios at L' and M from what is observed. In Figure 9 we plot the probability contours for our model in the $m, M$ plane. The red point is the case for microlensing only, no millilensing. The contours take on the shape of a very narrow band and there is a significant degeneracy along the line $m \times M = 1.79$. The extent of this degeneracy can be reduced only by the inclusion of longer wavelengths where the contribution of the accretion disk becomes insignificant. The small errors on our flux ratios make the contours very tight perpendicular to the $m\times M=1.79$ line. Changing the power law slope we used for the accretion disk moves the contours along the line $m\times M=1.79$ where steepening the power law index (bluer) will move the contours up to the left, and vice verse. 

The contribution for HS 0810 from the accretion disk in our simple model continues to be a factor even at longer wavelengths \citep[see][]{slus13}. Our model predicts a flux ratio of B/A = 0.9/1 will not occur until wavelengths longer than $\sim 10\micron$. The FWHM of the LBT at this wavelength is $0.3\arcsec$, but with a high signal-to-noise observation it is still possible to extract an accurate flux ratio given the very stable telescope PSF at this wavelength and the known separation of the two point sources.  The source brightness at $10~\micron$ is $\sim 30$ mJy,  which is sufficient given LBTI’s high sensitivity at infrared wavelengths. Although the upcoming JWST will be able to image much deeper at thermal MIR wavelengths than ground based telescopes such as the LBT, with a smaller 6 meter primary it will be difficult to resolve adjacent images in quad systems such at HS 0810. \cite{davi19} report detection of all four images at a wavelength of 2.1mm in the continuum and in CO line emission using ALMA. At this long wavelength the emitting region is likely considerably larger than the hot, inner dusty torus we observe in the MIR, and the images may be insensitive to millilensing on scales smaller than a few parsec, but capable of detecting substructure on larger scales. MIR imaging on the next generation of very large, ground based  telescopes equipped with an infrared AO system presents a future opportunity to further constrain the presence of millilensing due to sub-structure in dark matter halos. Alternative to long wavelength imaging, observations of emission from the quasar narrow line region, also unaffected by microlensing, shows promise \citep[e.g.][]{nier19}, but also requires high angular resolution.

\section{Conclusions}

We have imaged the lensed quasar system HS 0810+2554 at wavelengths spanning $2-5\micron$ with LMIRCam on the LBTI using the adaptive optics system. Flux ratios were computed for all four images in each filter. The two brightest images are $0.187 \arcsec$ apart, and have very high signal to noise at all wavelengths. At the shorter wavelength of $2.16\micron$ (rest wavelength $0.86\micron$) the flux is entirely consistent with flux from only the sub-parsec accretion disk of the quasar. At the two longer wavelengths, the flux is a combination of flux from the accretion disk and flux from a larger (few pc) dusty torus. A simple model employing multiplicative factors for image B due to stellar microlensing $(m)$ and sub-structure millilensing $(M)$ was presented. The result is tightly constrained to the product $m\times M=1.79$. The 60\% probability contour for this product stretches from $m= 2.6$, $M = 0.69$ to $m= 1.79$, $M = 1.0$, where the later is consistent with microlensing only. Note that we have not ruled out the presence of millilensing, but have shown that the microlensing only case is within the 60\% probability contour.

\section {Acknowledgments}

This publication makes use of data products from the Wide-field Infrared Survey Explorer, which is a joint project of the University of California, Los Angeles, and the Jet Propulsion Laboratory/California Institute of Technology, funded by the National Aeronautics and Space Administration.

This publication makes use of data products from the Two Micron All Sky Survey, which is a joint project of the University of Massachusetts and the Infrared Processing and Analysis Center/California Institute of Technology, funded by the National Aeronautics and Space Administration and the National Science Foundation.

We thank Gerd Weigelt and Charles (Chick) Woodward for providing standard star observations using LMIRCam on different dates.

\begin{figure}
\epsscale{1.0}
\plotone{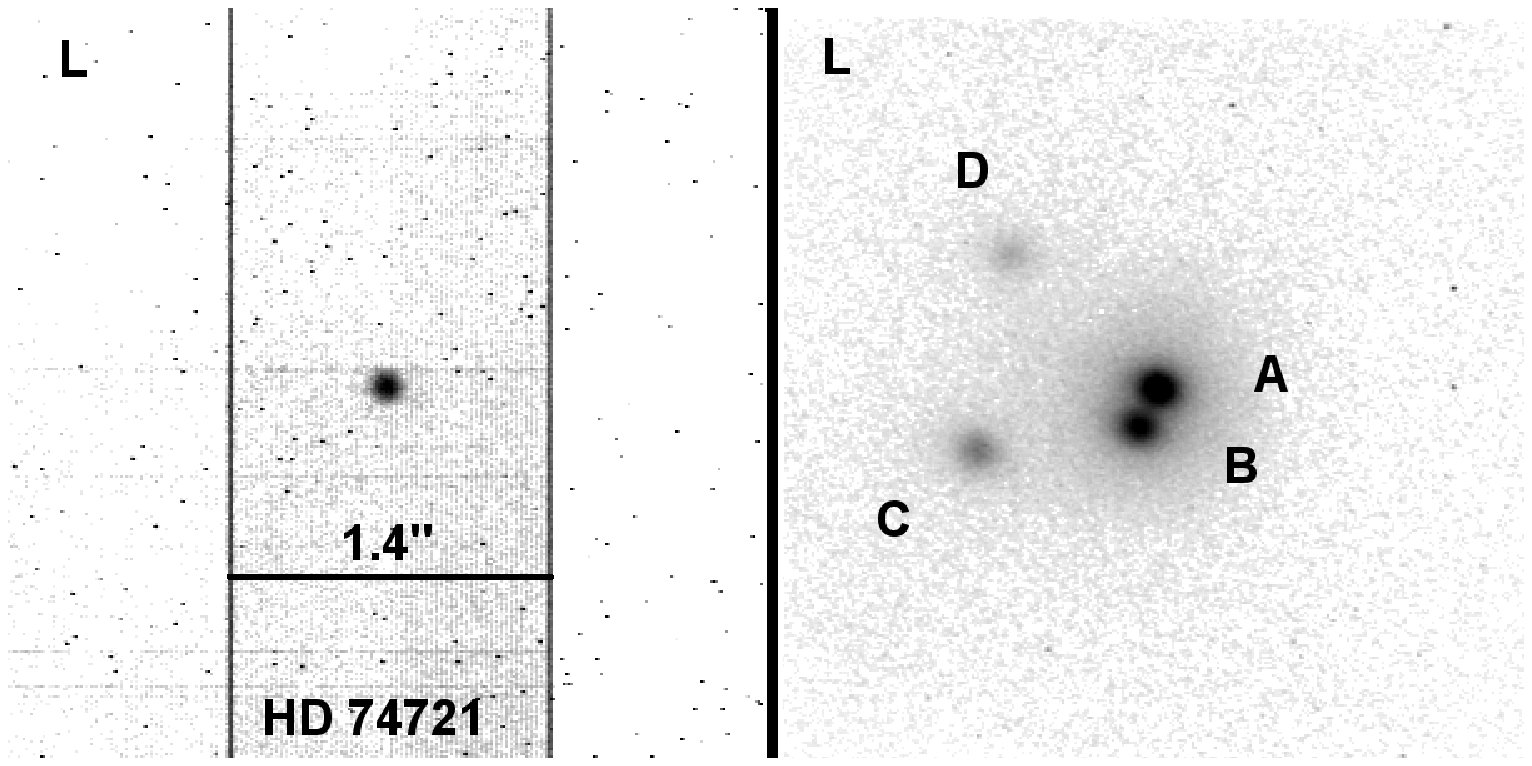}
\caption{Gray scale images of a standard star and the HS 0810 quad system in the L' band on 2017 February 17. The image of the star is only partially processed. The image of HS 0810 is fully processed and all four images of the source quasar are easily seen.}
\label{grey_std}
\end{figure}

\begin{figure}
\epsscale{1.0}
\plotone{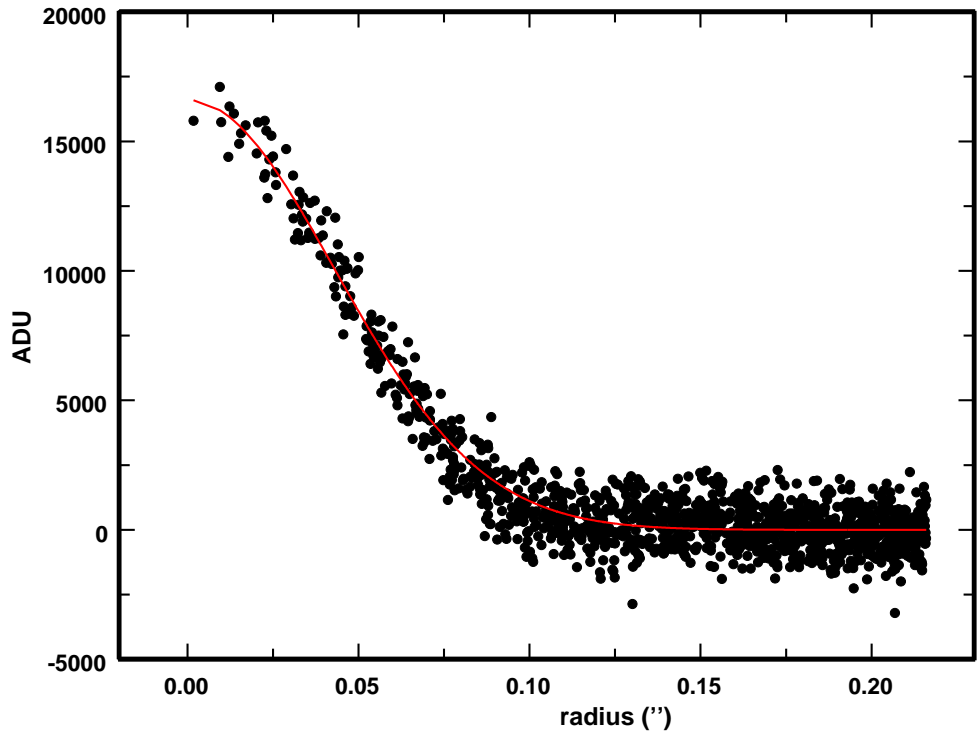}
\caption{Radial profile of the standard star shown in Figure 1. The profile is consistent with a Gaussian with a FWHM of $0.10\arcsec$.}
\label{psf}
\end{figure}

\begin{figure}
\epsscale{1.0}
\plotone{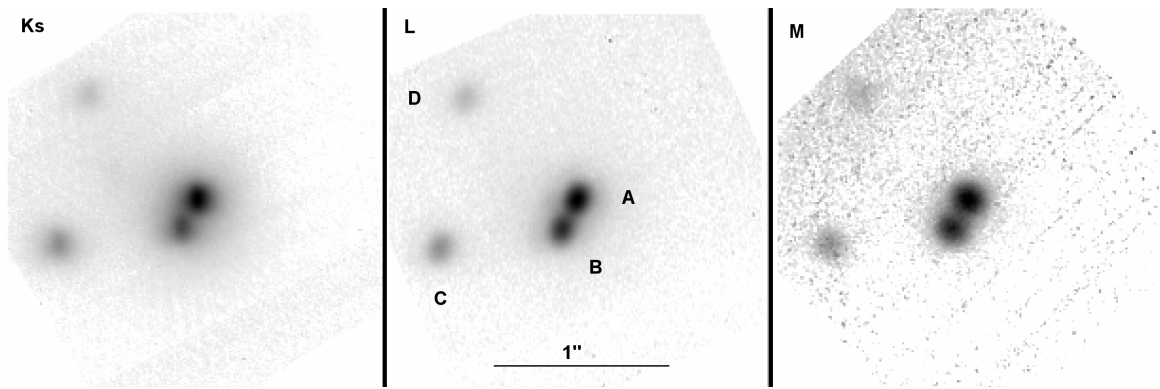}
\caption{Gray scale images of the HS 0810 Quad system at the three observed wavelengths. The lensing galaxy is just detectable at Ks left of center in the image.}
\label{grey}
\end{figure}

\begin{figure}
\epsscale{1.0}
\plotone{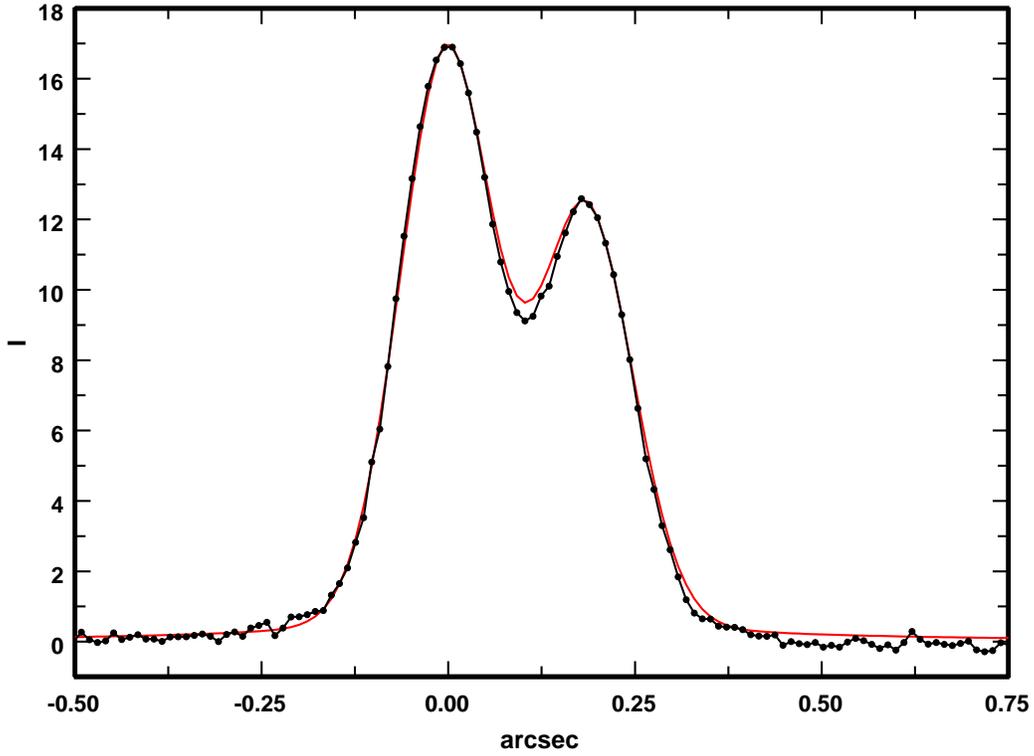}
\caption{Intensity (in ADUs) cut along an axis through the centers of images A and B at L' plotted as a black line. The model fit using two Voight functions is shown in red. The separation of the two peaks in the model is $0.187
\arcsec \pm 0.003\arcsec$}
\label{cut}
\end{figure}

\begin{figure}
\epsscale{1.0}
\plotone{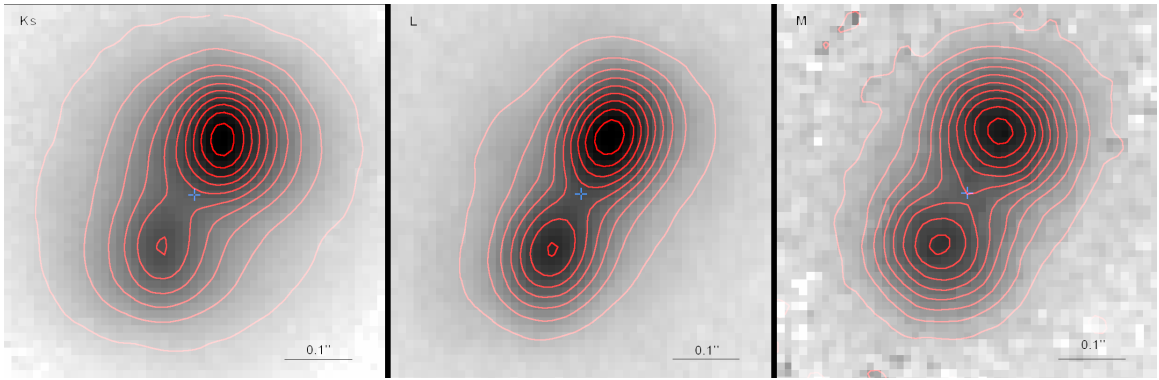}
\caption{Contours for images A and B arbitrarily scaled to produce 11 contours from sky to peak brightness. Note the clear trend towards more equal brightness with increasing wavelength.}
\label{contour}
\end{figure}

\begin{figure}
\epsscale{1.0}
\plotone{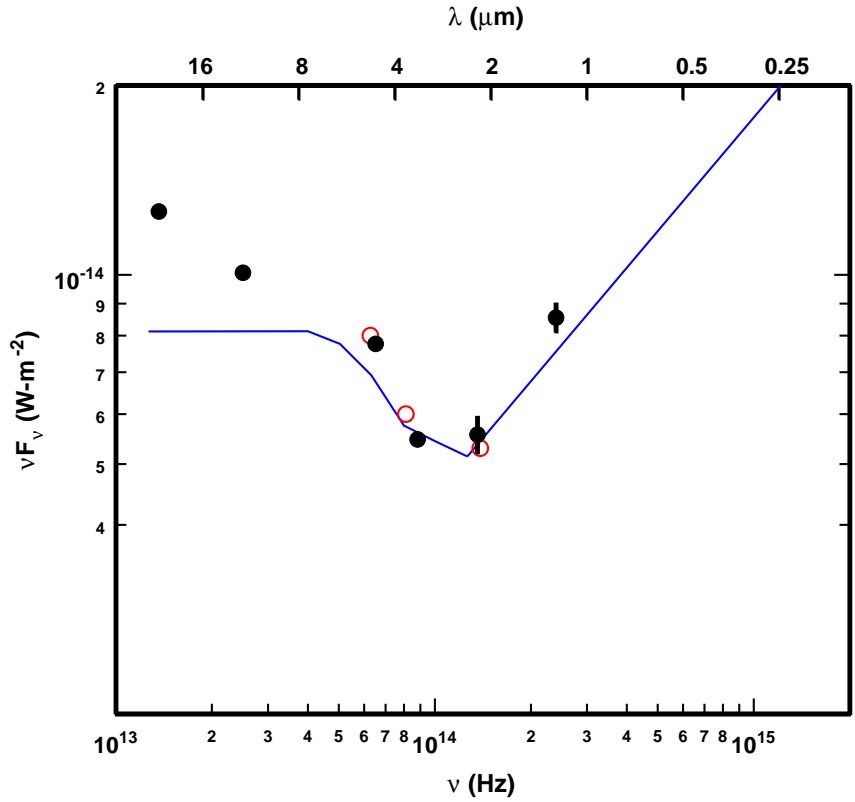}
\caption{Filled Circles: the spectral energy distribution of HS 0810 taken from the literature (see text). Blue Line: the mean quasar SED from \cite{kraw15}. Red open circles: combined flux in our filters scaled vertically to match the trend between the WISE W1 and W2 filters. The vertical bars indicate the range in fluxes between 2MASS and UKIDSS J and K band observations. The H band $(1.65~\micron)$flux is not plotted due to strong contamination by H$\alpha$.}
\label{SED}
\end{figure}

\begin{figure}
\epsscale{1.0}
\plotone{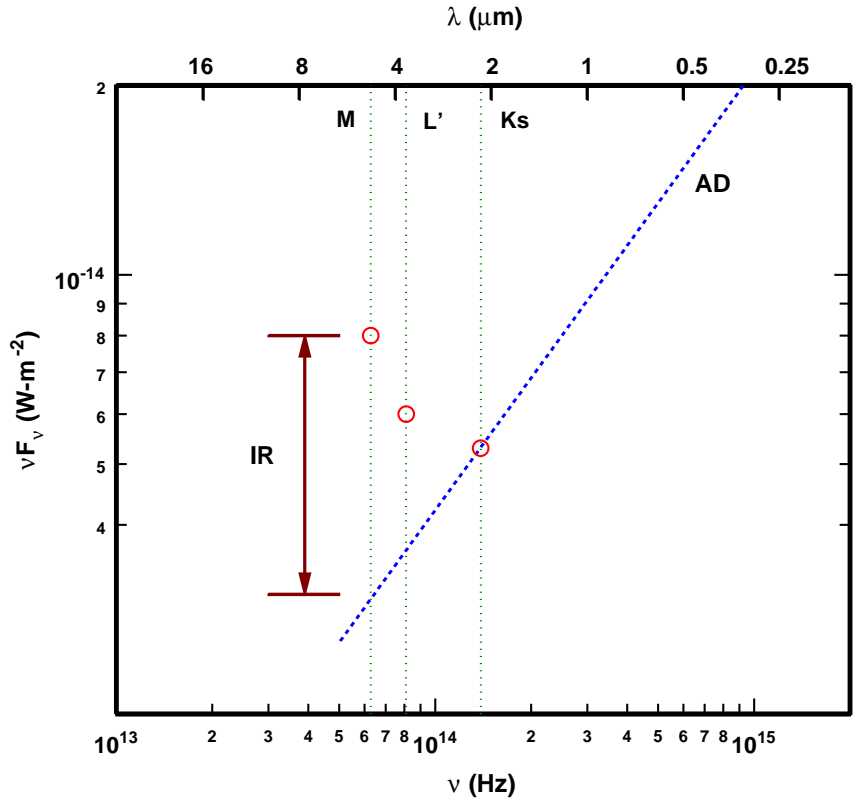}
\caption{Similar to Figure 6 except except a power law fit (dashed blue line) to the short wavelength portion is used to extrapolate the flux contribution from the physically small accretion disk to longer wavelengths. This power law fit is slightly steeper than the mean power law index shown in Figure 6 (see text). This yields the `AD' terms in Eq. 1. The vertical dashed lines indicate the location of our three filter bandpasses, and the difference between the accretion disk SED and the observed flux gives the `IR' terms in Eq. 1.}
\label{SED2}
\end{figure}

\begin{figure}
\epsscale{1.0}
\plotone{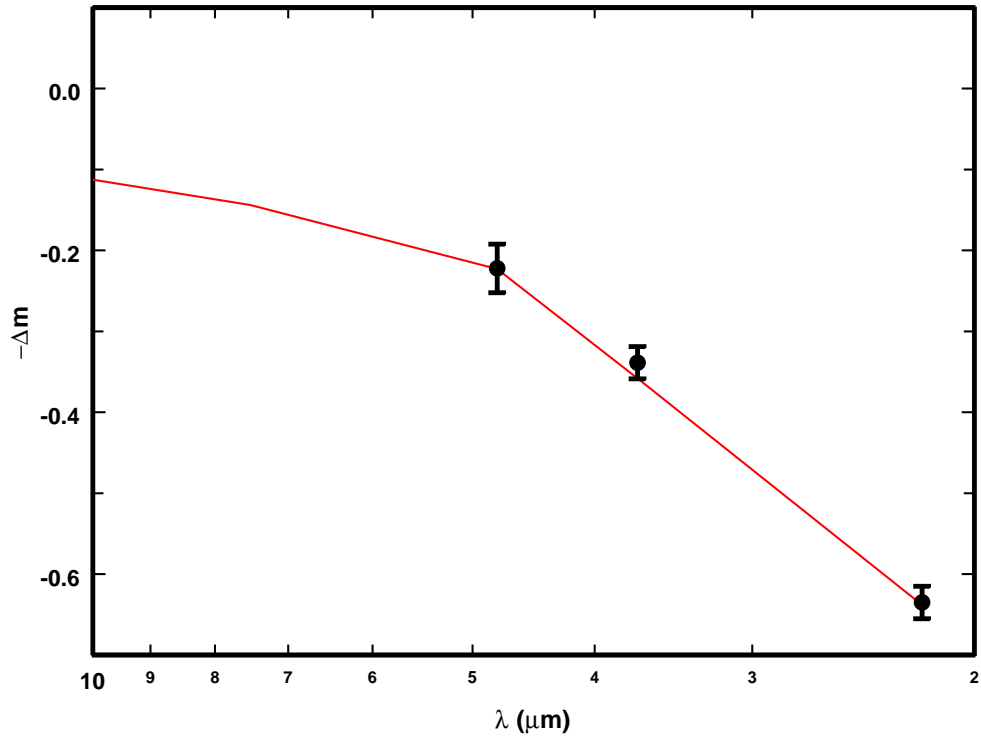}
\caption{Magnitude difference in brightness between images A and B as a function of observing wavelength. The red line is the model discussed in the text which has only microlensing, no millilensing $({\rm{M}}=1)$. }
\label{model}
\end{figure}

\begin{figure}
\epsscale{1.0}
\plotone{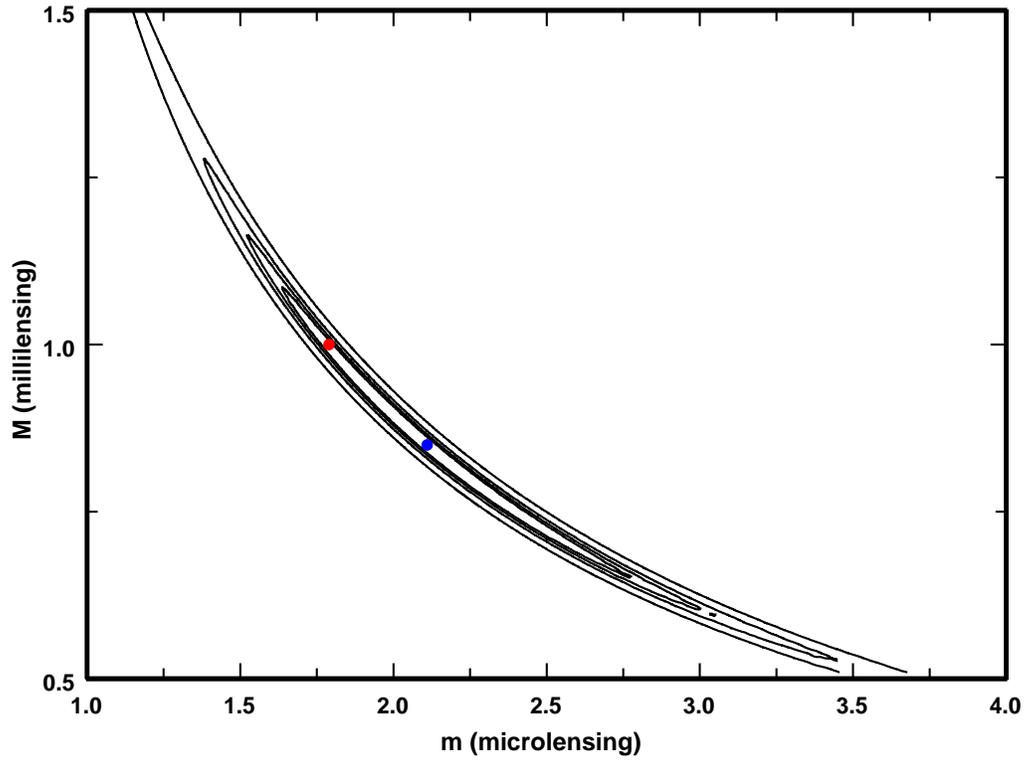}
\caption{Goodness of fit for the model as a function of microlensing (m) and millilensing (M). The outer contour is the 99\% confidence level with interior contours of 90, 80, and 70\%. The red point is the best fit model with microlensing only $(M = 1)$. The Blue point corresponds to the minimum in $\chi ^2$ with $m = 2.11$ and $M = 0.85$. }
\label{Chi2}
\end{figure}

\end{document}